Why a chloroplast needs its own genome tethered to the thylakoid membrane – Co-location for Redox Regulation

John F. Allen

Research Department of Genetics, Evolution and Environment
Darwin Building
University College London
Gower Street
London
United Kingdom
WC1E 6BT

E-mail: j.f.allen@ucl.ac.uk

**Abstract**

A chloroplast is a subcellular organelle of photosynthesis in plant and algal cells. A chloroplast's genome encodes proteins of the photosynthetic electron transport chain and ribosomal proteins required to express them. Chloroplast-encoded photosynthetic proteins are mostly intrinsic to the chloroplast's thylakoid membrane where they drive vectorial electron and proton transport. There they function in close contact with proteins whose precursors are encoded in the cell nucleus for cytosolic synthesis, subsequent processing, and import into the chloroplast. The protein complexes of photosynthetic electron transport thus contain subunits with one of two quite different sites of synthesis. If most chloroplast proteins result from expression of nuclear genes then why not all? What selective pressure accounts for the persistence of the chloroplast genome? One proposal is that photosynthetic electron transport itself governs expression of genes for its own components: co-location of chloroplast genes with their gene products allows redox regulation of gene expression, thereby resulting in self-adjustment of protein stoichiometry in response to environmental change. This hypothesis posits Co-Location for Redox Regulation, termed "CoRR", as the primary reason for the retention of genomes in both photosynthetic chloroplasts and respiring mitochondria. I propose that redox regulation affects all stages of chloroplast gene expression and that this integrated control is mediated by a chloroplast mesosome or nucleoid – a structure that tethers chloroplast DNA to the thylakoid.

## Main Text

Only a few percent of the thousand or so proteins in a chloroplast are encoded by chloroplast DNA. The remaining proteins are encoded by genes in the plant or algal cell nucleus, and their precursors are imported from the cytosol after synthesis on cytosolic ribosomes. What do chloroplast-encoded proteins have in common that has prohibited transfer of their genes to the nucleus? There are many proposals for the nature of selective pressures that maintain genes in organelles. These proposals tend to focus on the properties of individual proteins. Co-location for Redox Regulation, abbreviated CoRR, in contrast, focusses on the biochemical activity of the higher order structure within which chloroplast-encoded proteins function: photosynthetic electron transport. Chloroplasts exhibit the property of regulated gene expression. Chloroplast gene expression is not constant across the light-dark diel cycle, nor does it undergo random fluctuation. It is impacted by light, though no regulators of chloroplast gene expression are known to contain light-sensing prosthetic groups. Several independent lines of experimental evidence support the idea that chloroplast gene expression is under redox regulatory control (Allen 2015).

### Redox control of protein synthesis in isolated chloroplasts

The chloroplast contains a complete genetic system; from DNA replication and transcription to protein synthesis and assembly (Ellis 1981). One compelling line of evidence for redox control of chloroplast gene regulation is direct measurement of total chloroplast protein synthesis by incorporation of $^{35}$S-methionine into polypeptides synthesised within isolated pea chloroplasts (Allen et al. 1995a). The pattern of polypeptides newly synthesised in both thylakoids and stroma was unique to the redox conditions imposed during the incubation period of 45 minutes. These conditions included the presence or absence of specific chemical oxidants and reductants. Furthermore, distinct patterns of polypeptide labelling were obtained in light and darkness, with the light-dependent pattern being determined by the presence or absence of the site-specific inhibitors of photosynthetic electron transport DCMU and DBMIB. It was concluded that expression of chloroplast genes depends on the redox state of components of photosynthetic electron transport. Notably, light itself was not the stimulus impacting gene expression, it was the redox state of the chloroplast, which is altered by photosynthetic membrane function.

**Redox control of chloroplast transcription**

Chloroplasts evolved from cyanobacteria. Redox control of transcription for components of electron transport in prokaryotes entails the operation of either a two-component redox regulatory system (Allen 1993c) or a single-component redox activator or repressor (Allen 1996). In a series of experiments with chloroplasts isolated from mustard seedlings, synthesis of specific mRNAs took place in light of spectral composition favouring either photosystem I or photosystem II (Pfannschmidt et al. 1999a; Pfannschmidt et al. 1999b). Experiments with both mustard and pea chloroplasts indicate that plastoquinone redox state exerts control over transcription of genes for reaction centre apoproteins (Pfannschmidt et al. 1999a; Pfannschmidt et al. 1999b; Tullberg et al. 2000). These effects are functionally intelligible as responses correcting for imbalance in the stoichiometry of photosystem I and photosystem II (Allen and Pfannschmidt 2000; Pfannschmidt et al. 1999a). Initiation of transcription of genes for chloroplast proteins requires characteristically prokaryotic RNA polymerase subunits known as sigma factors (Macadlo et al. 2020). A number of factors exert regulatory control over photosynthesis and chloroplast function generally.  These include light intensity, nutrient status, and plastid developmental stage (Silverthorne and Ellis 1980).  Nevertheless, transcriptional initiation is proposed as a primary control point for plastid gene expression (Kayanja et al. 2021; Puthiyaveetil et al. 2010; Puthiyaveetil et al. 2021), and is likely to have been a conserved requirement in the ancestral cyanobacterium, in endosymbiosis, and in chloroplasts.

A number of redox enzymes are coupled to the chloroplast transcriptional apparatus (do Prado et al. 2024) that becomes tethered to the thylakoid membrane when transcription is active (Palomar et al. 2024).

**Translation and protein folding**

Folding of polypeptides occurs during synthesis, not after it, and begins within the exit tunnel of chloroplast ribosomes (Ries et al. 2020).  This folding is mediated by a chain of molecular chaperones (Ries et al. 2025) without which non-functional protein-protein interactions take place (Ellis 2004).  Early, guided protein folding may be of special importance for intrinsic membrane proteins whose hydrophobic side chains will otherwise promote protein aggregation. In addition, where the functional partner of the nascent membrane protein is itself a component of a pre-existing membrane protein complex then the ribosome engaged in translation becomes bound to the thylakoid membrane (Palomar et al. 2024).  Cryo-electron tomography of isolated, intact spinach chloroplasts reveals particles identified as ribosomes within 15 nm of the stroma-exposed "top" surface of thylakoid stacks (Wietrzynski et al. 2025).

**Membrane insertion during assembly of photosynthetic protein complexes.**

Electron flow through photosystem II and photosystem I needs to be stoichiometrically balanced. Chloroplast gene regulation helps to ensure physiological electron flow in the photosynthetic membrane.

The "D1" protein of the reaction centre of photosystem II becomes rapidly broken down and re-synthesised during a photosystem II repair cycle (Komenda et al. 2024; Li et al. 2024; Su et al. 2024). This cycle requires insertion of the membrane-intrinsic and hydrophobic D1 protein together with a post-translational cleavage of an amino-terminal segment of its precursor (Cheng et al. 2025). In cyanobacteria, the D1 precursor accumulates at thylakoid convergence zones (Ostermeier et al. 2025). Without exception D1 is always encoded in the genome of photosynthetically active plastids.

For photosystem I, in addition to the hydrophobic nature of membrane-intrinsic reaction centre proteins, land plant chloroplasts contain a thylakoid-bound photosystem I assembly factor, and assembly begins with the co-translational insertion of PsaA and PsaB into the thylakoid membrane to form the heterodimer of PsaA and PsaB in the reaction centre of photosystem I (Zhang et al. 2024).

**Tethering the genome to the thylakoid: from DNA transcription to assembly of photosynthetic membrane protein complexes**

Co-transcriptional translation is seen in prokaryotic systems (Qureshi and Duss 2025; Wang et al. 2020). I suggest that the chloroplast, derived from a cyanobacterium (Martin et al. 2002; Raven and Allen 2003), is no exception. If folding and insertion of nascent polypeptides accompanies the earliest stages of translation (Marino et al. 2016), and if translation occurs on ribosomes connected physically to the plastid-encoded RNA polymerase (do Prado et al. 2024), then a structural connection must exist between the thylakoid and chloroplast DNA. Co-transcriptional translation and co-translational assembly together require tethering between the chloroplast genome and the photosynthetic apparatus. This tethering is outlined schematically in Figure 1.

**A chloroplast mesosome?**

A mesosome is a point of attachment of DNA to a cell membrane in prokaryotes (Reusch and Burger 1973; Ryter 1968). Mesosomes were first observed by transmission electron microscopy of heavy-metal stained cells, raising the possibility that they were artifacts of sample preparation (Silva et al. 1976).  In early studies there was no obvious reason for a physical connection between a biological membrane and DNA, and several suggestions were made regarding their possible significance (Moyer 1979; Ryter 1968).  These suggestions tended to focus on cell division (Jacob et al. 1963; Jacob et al. 1966).  Noting a possible connection of the mesosome with redox reactions of the bacterial respiratory chain, Reusch and Burger wrote as follows: "Earlier expectations that the mesosome might represent the mitochondrion of the bacterial cell are not likely to be fulfilled. It would be of interest to study any possible relationship between the photosynthetic apparatus and mesosomes in photosynthetic bacteria" (Reusch and Burger 1973).

**The chloroplast nucleoid**

A nucleoid was originally thought to be a bacterial equivalent of the eukaryotic cell nucleus, also with a focus on its possible role in cell division (Iterson et al. 1975).  Today the idea of a chloroplast mesosome seems to have fallen out of favour, while there are active studies and a detailed composition (Ahrens et al. 2025; Pfalz and Pfannschmidt 2013) of what is agreed to be a chloroplast nucleoid: "In bacteria the genome is organized as a huge protein/DNA complex that is not separated from the cytosol by a nuclear double membrane as in eukaryotes. It is thus accessible to the transcription and translation machineries at the same time. The plastome is organized in a similar manner…" (Ahrens et al. 2025). Chloroplast transcription can be studied in a transcriptionally active chromosome, or TAC, that contains DNA, RNA, and around 35 proteins; at least five of which are predicted to be membrane-bound (Ahrens et al. 2025).

Protein and RNA structural features are seen in a model derived from cryo-electron microscopy of the complete spinach chloroplast ribosome at 3.0 Å resolution (Perez Boerema et al. 2018) (reviewed in (Allen 2018)).

**mRNA targeting**

Cyanobacterial mRNAs that encode core photosynthetic membrane proteins are themselves attached to the thylakoid by means of membrane-associated RNA-binding proteins (Hess et al. 2025). Co-translational folding and assembly indicate co-translational insertion of nascent polypeptide chains, or "transertion": "a system for membrane protein production in which transcription of an mRNA and co-translational insertion of its translation product occur simultaneously. It requires the genomic locus encoding the protein to be in close proximity to the target membrane" (Hess et al. 2025) . It is proposed (Hess et al. 2025) that that transertion was maintained throughout the endosymbiotic incorporation of a cyanobacterium into the cytoplasm of a eukaryote, and that this process required the maintenance of an internal genome to supply the mRNAs required. Co-transcriptional translation together with co-translational insertion of precursor polypeptides requires tethering between the chloroplast genome and the photosynthetic apparatus, as outlined schematically in Figure 1. Transertion (Hess et al. 2025) provides direct evidence of physical interaction between regulated genome expression and the regulated function of the photosynthetic electron transport chain, the salient element of the CoRR hypothesis.

**Integrated redox control of multiple stages of chloroplast gene expression**

The CoRR hypothesis can be illustrated as the need for DNA internal to the chloroplast in order to permit redox regulatory control of transcription (Allen 1993a, b, 2003, 2015, 2017). There are independent lines of evidence indicating the existence of such control (Pfannschmidt et al. 1999a; Pfannschmidt et al. 1999b; Puthiyaveetil et al. 2013; Puthiyaveetil et al. 2010). Indeed transcriptional initiation may be a primary control point in chloroplast gene expression (Puthiyaveetil et al. 2021). Nevertheless there are clear indications that every stage of chloroplast gene expression is under redox control, and that none, on its own, is sufficient to explain integrated photosynthetic redox signalling (Allen et al. 1995b). Post-transcriptional regulation of plastid gene expression may occur by means of RNA processing that is under the control of nuclear genes (Barkan 2011; Hu et al. 2025). Ribosomal DNA footprinting reveals multiple epistatic effects of mutations in chloroplast DNA (Zoschke et al. 2013). *psbA* translation is light-regulated by means of a factor or factors involved in the photosystem II repair cycle (Chotewutmontri and Barkan 2018, 2020).

Other theories for organelle genome retention emphasise the need for light-regulated gene expression, including Control by Epistasy of Synthesis (CES) (Ghandour et al. 2025; Wietrzynski et al. 2021; Wollman et al. 1999) in which synthesis of chloroplast-encoded photosynthetic subunits is regulated by the assembly state of the complexes in which they function. In angiosperms results from ribosome profiling and RNAseq indicate that *psbA* is the only gene in a mature chloroplast whose expression is regulated by light, with this regulation occurring at the level of translation (Chotewutmontri et al. 2025). Given an apparent switch between transcription and translation during development (Silverthorne and Ellis 1980), this conclusion may be consistent with the observed control of *psbA* transcription in seedlings of mustard (Pfannschmidt et al. 1999b) and pea (Tullberg et al. 2000). In dinoflagellates there is also CO-location for Control of Assembly (COCOA) (Howe and Barbrook 2024), a process that clearly depends on gene retention. These proposals are not inconsistent with the CoRR hypothesis, and may be seen as special cases of CoRR's central postulate: the non-negotiable evolutionary requirement for regulated function of bioenergetic membranes as the selective pressure that maintains genes (hence DNA) in organelles. In particular, the redox state of the thylakoid plastoquinone pool may be decisive for numerous regulatory responses to environmental stress (Wilson et al. 2006).

**Inference from comparison of organellar genomes**

Evolutionary retention of genes in organelles is correlated with hydrophobicity of protein products of organellar gene expression (Giannakis et al. 2022); with GC content; and with "the centrality of a protein product in its functional complex" (Giannakis et al. 2024). Hydrophobicity alone has been proposed as the decisive factor for gene retention, particularly for mitochondria (von Heijne 1986), with subsequent enquiries then focussing on possible barriers to import of protein precursors from the cytosol (Björkholm et al. 2017). In such studies there seems to be a tacit assumption that endosymbiotic gene transfer (Ku et al. 2015) – migration of genes from an endosymbiont to the eukaryotic cell nucleus (Huang et al. 2003; Martin 2003) – is inevitable, even if slow (Palmer 1997), and thus reasons will be found for barriers to this migration in the case of specific genes retained by the cell nucleus. This and other proposals for retention of organellar genomes have been reviewed (Allen 2003). Both chloroplasts and mitochondria have retained genes for proteins of their respective bioenergetic membranes (Maier et al. 2013). This remarkable convergence mat be the result of the same selective force—the need for balanced electron flow—acting upon independently-evolving units of physiological function: the photosynthetic and respiratory electron transport chains. By contrast, there is no structural feature that is common to all chloroplast-encoded gene products. Instead, conserved patterns of gene location are consistent with photosynthetic redox chemistry exerting gene regulatory control over its own rate-limiting steps (Allen et al. 2011). Chloroplast DNA carries genes whose expression is placed under this control. The CoRR hypothesis proposes that gene location is itself the result of natural selection, and predicts that genes will never disappear completely from photosynthetic organelles. This view seems to be especially clear in the case of chloroplasts (Allen et al. 2005) where photosynthetic redox control of chloroplast gene expression (Allen 1993b) is now well established (Allen 2015, 2017) .

**Acknowledgement**

I thank William F. Martin, Michael J. Russell, and Elinor P. Thompson for comments on the manuscript.

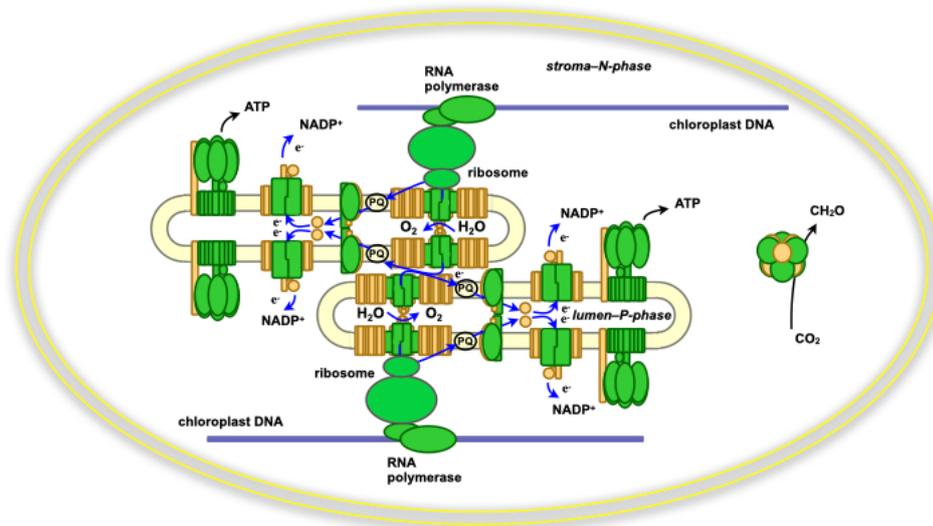

**Figure 1.** A schematic outline of chloroplast DNA tethered to two, stacked thylakoids. Green shapes represent proteins encoded in chloroplast DNA; brown shapes represent proteins encoded in the cell nucleus. Blue arrows represent the path of photosynthetic electron transport. PQ; plastoquinone.